\documentclass{article}

\usepackage{amssymb}
\usepackage{subcaption}
\usepackage{amsthm}
\usepackage{amsmath}
\usepackage{natbib}
\usepackage{afterpage}

\usepackage[colorlinks,citecolor=blue,urlcolor=blue,filecolor=blue,backref=page]{hyperref}
\usepackage{graphicx}

\newtheorem{example}{Example}
\newtheorem{definition}{Definition}  % Define the "definition" environment

\title{Bayesian Graphs of Intelligent Causation}
\author{Preetha Ramiah\footnote{Department of Statistics, University of Warwick, UK.}, James Q. Smith\footnote{Department of Statistics, University of Warwick, UK and The Alan Turing Institute, UK.}, Oliver Bunnin\footnote{Natwest Markets plc, UK.}, \\ Silvia Liverani\footnote{School of Mathematical Sciences, Queen Mary University of London, UK.}, Jamie Addison \footnote{Dstl, UK}, Annabel Whipp \footnote{Dstl, UK}}
\date{}

\begin{document}

\maketitle

\begin{abstract}
Probabilistic Graphical Bayesian models of causation have continued to impact on strategic analyses designed to help evaluate the efficacy of different interventions on systems. However, the standard causal algebras upon which these inferences are based typically assume that the intervened population does not react
intelligently to frustrate an intervention. In an adversarial setting
this is rarely an appropriate assumption. In this paper, we extend an established Bayesian methodology called Adversarial Risk Analysis to apply it to settings that can legitimately be designated as causal in this graphical sense. To embed this technology we first need to generalize the concept of a causal graph. We then proceed to demonstrate how the predicable intelligent reactions of adversaries to circumvent an intervention when they hear about it can be systematically modelled within such graphical frameworks, importing these recent developments from Bayesian game theory. The new methodologies and supporting protocols are illustrated through applications associated with an adversary attempting to infiltrate a friendly state.
\end{abstract}

%% ** Keywords **
%\begin{keyword}%[class=MSC]
%Causality, Graphical Models, Adversarial Risk, Bayesian Networks, Chain Event Graphs, Causal Interventions, Structural Reasoning 
%\kwd{}
%\kwd[]{}
%\end{keyword}

%% ** Mainmatter **
\section{Introduction}

Causal models are variously defined. In this paper, we will restrict ourselves
to discuss only Bayesian causal graphical models whose graphical semantic have
been customised to a particular application. Such causal graphical models
embed a collection of structural hypotheses about probabilistic predictions
that are assumed to hold not just for an \emph{idle} system - i.e. one which is
not controlled - but also to hold for predictions of the effects of the contemplated
classes of interventions. This type of causal analysis is already being used
to support Bayesian strategic analyses to help investigate the uncertain
future impact on populations or infrastructure of various types of
intervention. Examples of this are the well known graphical framework, the Causal 
Bayesian Network (CBN) \citep{Pearl2000,Spirtesetal1993}, and more recently the flow network
\citep{smith2007causal}, the regulatory graph \citep{liverani2016bayesian} and the chain event graph \citep{ThwaitesEtAl2010}. It appears
that the choice of an appropriate graphical model is often critical to
represent those qualitative features that can be assumed invariant under the
class of interventions we might contemplate making. 

Although Bayesian causal modelling is now well advanced, the underlying
algebras they use typically assume that the intervened population cannot be
expected to intelligently attempt to mitigate the effects of any contemplated
intervention. Inspired by the early work of \cite{hill1965environment}, this
paper develops new Bayesian tools that extend such causal analyses so that
the effects of potential intelligent resistance are taken into account. One
key additional causal assumption we make here is that the defender believes that their chosen graphical framework will not only be invariant to their
own decisions, but also hold true for other intelligent people such as the
adversary. Bayesian causal analyses are thus extended to support strategic
analysis where they can use a bespoke causal graph to put themselves in the
shoes of an adversary and to systematically take account of the potential
intelligent reactions a defensive intervention might
induce. We adapt and apply the now well-developed Adversarial Risk Analysis
(ARA) methodology to  scenarios which can plausibly be considered causal in this
sense. Although our methodology applies more generally, to avoid some of the
more abstruse game theoretic issues that would be otherwise necessary to
discuss for simplicity and clarity, we here focus only on applications
where a defender faces a single adversary in a simple defend - attack game \citep{banks2015,naveiro2022}. 

One standard idea we import from game theory associated with gathering intelligence, recently discussed in \cite{riosinsua2023}, is the
concept that it is important that the
defender models the adversary's intent (goals), capability and knowledge to allow for intelligent reactions. This then enables the defender to model not only what the adversary typically does, but also how they might react to the defender's interventions. Within our coloured
graphical framework these additional, often latent, features are captured by
embellishing a graphical model of the idle system with further latent nodes edges and
colours. The model of the idle system can then be viewed as a margin of this
new enlarged model from which predictive distributions under control - after
folding in likely adversarial responses - can be calculated. 

When such a causal graphical framework can be introduced into an ARA analysis
we observe that the invariance assumptions it implies can often make the
otherwise sensitive task of the defender double guessing their adversary's
beliefs much more straightforward than it would otherwise be. This is because
many features of the problem can be assumed to be shared by the two players. This includes the modularisation of the system induced by the shared customised
graphical framework. This means that the defender needs to only contemplate their
adversary's reaction since it affects certain specific conditional densities or
mass functions. We illustrate a utilisation of this modularisation with a
simple example, where an enemy agent plans to infiltrate an agent into a friendly state. So we argue that there is an exciting potential symbiosis
between ARA and causal analyses within the Bayesian paradigm which could be
further exploited in seriously complex adversarial environments. 

In the next section, we present, for the first time, a definition of causal algebras
that can be applied to general classes of causal graphs, rather than just to BNs,
and illustrate how this definition applies to standard graphical causal models with
intelligent reaction to an intervention. In Section 3, we then modify ARA methodologies to develop new technologies to show how our generic definition
provides an appropriate inferential framework to model causal processes that
explicitly acknowledge the intelligent reactions of an adversary. We then
proceed to illustrate how this technology can be applied within a Bayesian
defender's strategic analysis of how best to defend against various forms of threatened enemy infiltration. We conclude the paper with a short discussion about
how these new techniques are being used by practitioners to study more complex
adversarial domains and future challenges, and how new causal discovery
algorithms can be developed for adversarial domains to provide practical
decision support. 

\section{Generic Classes of Causal Algebras}

%\subsection{Some formal definitions}

We first define a generic graphically-based causal algebra, rich enough not only to encompass Bayesian models describing the
progress of plots that are likely to be perpetrated by an adversary $A$, but
also the intelligent responses that such interventions might induce. So
suppose a defender $D$ plans to perform a strategic or forensic analysis to
compare the efficacy of a set of possible policies - here called \emph{%
interventions} $d\in \mathbb{D}$ where $\mathbb{D}$ includes doing nothing $%
d_{\emptyset }$. These decisions are designed to counteract the impacts of the malign
activities of an adversary $A$. The defender's utility function, $U(d,\boldsymbol{e})$%
, is a function of the interventions $d$ over a vector $%
\boldsymbol{e}$ of utility attributes, or \emph{effects} in causal terminology, which fully capture the consequences of each potential intervention $d$.

The defender\ now selects a (possibly coloured) graphical model class $%
\mathbb{G}$ - with a particular semantic $\mathcal{S}_{\mathbb{G}}$ - which
they believe will be valid for all $\mathcal{G\in }\mathbb{G}$ regardless of
the topology of the graph $\mathcal{G}$ and the colouring of its vertices
and edges, and customised to the modelled domain and $D$'s reasoning. A
critical property of this causal graphical model $\mathcal{G}$ is that it must be sufficiently refined to not only explain the idle process, but also the necessary probability
model of the effects of each $d \in \mathbb{D}$.

Under their chosen semantics $\mathcal{S}_{\mathbb{G}}$, the topology of
each $\mathcal{G}$ will depict a collection of causally
associated natural language statements about the underlying process
generating\ both the idle system and intervened system. Through $\mathcal{S}%
_{\mathbb{G}}$ these natural language statements are then embellished into
statements about a random vector constructed by the analyst, denoted here by $\boldsymbol{\xi }$. The
distribution of effects $\boldsymbol{e}$ can then be calculated as a margin
of $D$'s posterior densities $p_{d}(\boldsymbol{\xi })$. 

To perform a Bayesian decision analysis, for each $d\in \mathbb{D}$, $D$
calculates their subjective expected utility scores of each contemplated
interventions $\left\{ \overline{U}(d):d\in \mathbb{D}\right\} $ where
\begin{equation}
\overline{U}(d)\triangleq \int U(d_{\emptyset },\boldsymbol{e})p_{d}(%
\boldsymbol{e})d\boldsymbol{e}.  \label{expu}
\end{equation}%
Within this graphical framework, we assume that there is a function $f_{%
\mathcal{G}}$ determined by $D$'s choice of $\mathcal{G}$ and its semantics 
$\mathcal{S}_{\mathbb{G}}$ such that   
\begin{equation}
p_{d\mathcal{G}}(\boldsymbol{\xi })=f_{\mathcal{G}}\left( p_{d_1\mathcal{G}}(%
\boldsymbol{\xi }),p_{d_2\mathcal{G}}(\boldsymbol{\xi }),\ldots ,p_{d_m%
\mathcal{G}}(\boldsymbol{\xi })\right)   \label{prob_zi}
\end{equation}%
where for each  $d\in \mathbb{D}$  
\begin{equation}
\mathcal{P}_{d\mathcal{G}}\triangleq \left\{ p_{d_1\mathcal{G}}(\boldsymbol{%
\xi }),p_{d_2\mathcal{G}}(\boldsymbol{\xi }),\ldots ,p_{d_m\mathcal{G}}(%
\boldsymbol{\xi })\right\} 
\end{equation}%
are an ordered sets of \emph{factors }that determine $p_{d\mathcal{G}}(%
\boldsymbol{\xi })$ for each $d\in \mathbb{D}$. Here we write $\mathcal{P}_{%
\mathcal{G}}\triangleq \mathcal{P}_{d_{\emptyset }\mathcal{G}}$ and $p_{%
\mathcal{G}}(\boldsymbol{\xi })\triangleq p_{d_{\emptyset }\mathcal{G}}(%
\boldsymbol{\xi })$.

Clearly the relationship between $\mathcal{P}_{\mathcal{G}}$ and $\mathcal{P}%
_{d\mathcal{G}}$ for $d\neq d_{\emptyset }$ is critical to this analysis. One
of the key properties to infer the likely impact of each potential intervention from information embedded in an idle system, as for any good causal model, is to specify those features of the model which can be conjectured to be invariant under such actions \citep{PetersRSS}. In the context of graphical based causation this translates
into two conditions. The first is that the causal graph $\mathcal{G}$ and the
function $f_{\mathcal{G}}$ defined in Equation (\ref{prob_zi}) will continue to hold - i.e. be invariant to - the application of any contemplated intervention $d\in \mathbb{D}$. The second is that if $%
\mathbb{G}$ and $\mathcal{G}$ are well chosen then the effect of each $d\in 
\mathbb{D}$ should be local in the sense that most of the factors in $%
\mathcal{P}_{d\mathcal{G}}$ will be shared with those in $\mathcal{P}_{%
\mathcal{G}}$. 

We partition $\mathcal{P}_{d%
\mathcal{G}}$ into the set of factors $\mathcal{P}_{d\mathcal{G}}^{0}$
shared with the idle factors $\mathcal{P}_{\mathcal{G}}$ and its complement $%
\mathcal{P}_{d\mathcal{G}}^{1}$.
\[
\mathcal{P}_{d\mathcal{G}} = \mathcal{P}_{d\mathcal{G}}^{0} \cup \mathcal{P}_{d\mathcal{G}}^{1}
\]
where $\mathcal{P}_{d\mathcal{G}}^{0} = \mathcal{P}_{d\mathcal{G}} \cap \mathcal{P}_{\mathcal{G}}$ and $\mathcal{P}_{d\mathcal{G}}^{1} = \mathcal{P}_{d\mathcal{G}} - \mathcal{P}_{d\mathcal{G}}^{0}$.

The probability model of the idle system
in a given class of models $\left( \mathbb{G},\mathcal{S}_{\mathbb{G}%
}\right) $ is fully specified by $\left( \mathcal{G},f_{\mathcal{G}},%
\mathcal{P}_{\mathcal{G}}\right) .$ Therefore because of the invariance of $%
\left( \mathcal{G},f_{\mathcal{G}}\right) $ - a property that elevates the
idle model to a causal one - by specifying $\left\{ \mathcal{P}_{d\mathcal{G}%
}^{1}:d\in \mathbb{D}\right\}$, $D$ can calculate all the scores they need
for their analyses of the consequences of each $d\in \mathbb{D}$. These
basic principles underpin not only causal systems framed by a Bayesian Network (BN), but many
others. These are used to define the generic graphical causal frameworks
discussed below. 

The topology of the selected (possibly coloured) graph $\mathcal{G}$ and its
underlying semantics $\mathcal{S}_{\mathbb{G}}$ - customised to the
adversarial setting in focus - therefore together have four critical roles
within such a Bayesian causal decision analysis:
\begin{enumerate}
\item $\mathcal{G}$ will provide an \emph{interface} between $D$'s natural
language causal description of not only what is currently happening, but what
might happen if any of the contemplated interventions $d\in \mathbb{D}$ were
to be enacted. It will be demonstrated below that behavioural models
rational adversaries require an \emph{explanatory} $\mathcal{G}$ that is
considerably more refined than in other settings.
\item $\mathcal{G}$ determines what \emph{factors} $\left\{ \mathcal{P}_{d%
\mathcal{G}}:d\in \mathbb{D}\right\} $ are needed to embellish $\mathcal{G}$
into a full probability model. 
\item The semantics $\mathcal{S}_{\mathbb{G}}$ of $\mathcal{G}$ will also
determine the \emph{function} $f_{\mathcal{G}}$ that is needed to combine the
factors $\mathcal{P}_{d\mathcal{G}}$ to derive $\left\{ p_{d\mathcal{G}}(%
\boldsymbol{\xi }):d\in \mathbb{D}\right\}$.
\item $\left( \mathcal{S}_{\mathbb{G}},\mathcal{G}\right) $ determine a
partition of $\mathcal{P}_{d\mathcal{G}}$ into two disjoint sets of factors $%
\mathcal{P}_{d\mathcal{G}}^{0}$ and $\mathcal{P}_{d\mathcal{G}}^{1}$
defining those factors that are invariant to a particular intervention $d\in 
\mathbb{D}$. 
\end{enumerate}

Let the density $p_{\mathcal{G}}(\boldsymbol{\xi })\in \mathbb{P}_{\mathcal{G%
}}$ of the idle system be constructed from $\left( \mathcal{G},\mathcal{S}_{%
\mathbb{G}},\mathbb{P}_{\mathcal{G}}(f_{\mathcal{G}},\mathcal{P}_{\mathcal{G}%
}),\boldsymbol{\xi }\right) $ where all these terms are defined above.

\begin{definition}\label{def1}
Call a map $\Phi $ \emph{causal} with respect to a set of
interventions $\mathbb{D}$ if for every $d\in \mathbb{D}$ there is a
deterministic well defined map $\Phi $ such that 
\begin{eqnarray*}
\Phi  &:&\mathbb{P}_{\mathcal{G}}\times \left\{ \mathcal{P}_{d\mathcal{G}%
}^{1}:d\in \mathbb{D}\right\} \rightarrow \mathbb{P}_{\mathcal{G}} \\
&:&\left( p_{\mathcal{G}}(\boldsymbol{\xi }),\left( p_{d1\mathcal{G}}(%
\boldsymbol{\xi }),p_{d2\mathcal{G}}(\boldsymbol{\xi }),\ldots ,p_{dm%
\mathcal{G}}(\boldsymbol{\xi })\right) \right) \mapsto p_{d\mathcal{G}}(%
\boldsymbol{\xi })
\end{eqnarray*}%
where $p_{d\mathcal{G}}(\boldsymbol{\xi })$ is defined by Equation (\ref{prob_zi}). The
formulae $$\left\{ f_{\mathcal{G}}\left( p_{d1\mathcal{G}}(\boldsymbol{\xi }%
),p_{d2\mathcal{G}}(\boldsymbol{\xi }),\ldots ,p_{dm\mathcal{G}}(\boldsymbol{%
\xi })\right) :d\in \mathbb{D}\right\} $$ together constitute  the \emph{%
causal algebra} associated with $\Phi $. 
\end{definition}

Given $\mathcal{G}$, if the factors in $\mathcal{P}_{\mathcal{G}}$ were known
then by specifying $\mathcal{P}_{\mathcal{G}}^{1}$ the map $\Phi $ provides
the inputs needed for a subjective expected utility (SEU) analysis. However, in practice such factors
will be uncertain. Furthermore,\ certain components of $\boldsymbol{\xi }$
may be completely latent. Indeed we argue below that in an adversarial
setting this will almost always be the case. However, this challenge is not a
new one. For example there is now an enormous literature about when causal
effects are identifiable from a massive data set, but only from a given margin of the process \citep{Pearl2000,Spirtesetal1993}. 

From a Bayesian perspective, by introducing informative priors into the
analysis, we can still proceed formally in this setting to specify
the SEU scores needed for a strategic analysis. Moreover, the causal algebra and
its factorisation guides us in determining how we can calculate the
necessary scores numerically. So such issues become simply computational
ones, albeit sometimes challenging. %In this paper we only consider settings where such computations are straightforward.

Definition \ref{def1} is sufficiently general for us to perform formal causal
analyses in the adversarial settings we address in this paper where the Bayesian Network
might not be the most appropriate decision analytic graphical framework.

\subsection{Bayesian Networks}

There are now many classes $\mathbb{G}$ and semantics $\mathcal{S}%
_{\mathbb{G}}$ of graphical models $\mathcal{G}$
which express various collections of structural
conditional independence statements about a given set of random variables $%
\boldsymbol{X}=\left\{ X_{1},X_{2},\ldots X_{n}\right\}$. The first use of a
graphical class $\mathbb{G}$ to express causal relationships and be the
basis of a causal algebra was the Bayesian Network (BN) \citep{Pearl2000,Spirtesetal1993}. This class has proved to be ideal in applications where most of the
structural information that might be preserved under intervention is about
certain types of dependence relationships between a pre-specified set of
measurements $\boldsymbol{X}$.

The vertices of the uncoloured directed acyclic graph $\mathcal{G}$ within the class of
BNs $\mathbb{G}$ represent the random variables $\boldsymbol{X}=\left\{
X_{1},X_{2},\ldots X_{n}\right\} $. Here we use the notation that the
subvector $\boldsymbol{X}_{Q(i)}$ of $\boldsymbol{X}$ represents the parents
of $X_{i}$ in $\mathcal{G}$, for $i=1,2,\ldots ,n$, that is, the vertices in $%
\mathcal{G}$ which have a directed edge into $X_{i}$. In the notation we use above for generic causal models, $m=n$ since the explanatory variable $\boldsymbol{\xi }$ can
be simply identified with $\boldsymbol{X}$ and the number of factors
associated with conditional densities of each random variable.

For this class $\mathbb{G}$ the most studied class of interventions $d_{I(%
\widehat{\boldsymbol{x}})}\in \mathbb{D}$ are ones which force the
random variable $X_{i}$ to take the value $\widehat{x}_{i}$ for each $i\in
I\subseteq \left\{ 1,2,\ldots ,n\right\} $ with $\widehat{\boldsymbol{x}}%
\triangleq \left\{ \widehat{x}_{i}:i\in I\right\} $. Such interventions are
commonly referred to as ``doing'' $\left\{ X_{i}=\widehat{x}_{i}:i\in
I\right\} $. When the causal algebra we specify below is asserted to hold
for all $I\subseteq \left\{ 1,2,\ldots ,n\right\} $ the BN is called a \emph{%
causal BN}.

The class of BNs is one example of our more general definition and enjoys
all the properties we describe above.
\begin{enumerate}
\item The graph $\mathcal{G\in }\mathbb{G}$ embodies a collection of natural
language statements about the idle system of the form, for example, ``If I had available $%
{X}_{B}$ and ${X}_{C}$ to help forecast value of
measurement ${X}_{A}$ once I knew the value ${x}_{C}$
of ${X}_{C}$, the value ${x}_{B}$ of ${X}%
_{B}$ would be irrelevant to any forecasts about the (as yet unknown) value of $%
{X}_{A}$''. This is often written ${X}_{A}\amalg 
{X}_{B}|{X}_{C}$. The graph $\mathcal{G}$ represents a
particular type of collection of such statements within its semantics $%
\mathcal{S}_{\mathbb{G}}$. %For example the graph 
%\begin{equation*}
%\begin{array}{ccc}
%X_{2} & \rightarrow & X_{4} \\ 
%& \searrow &  \\ 
%X_{1} & \rightarrow & X_{3}%
%\end{array}%
%\end{equation*}%
%embodies the two non-trivial natural language statements, that $X_{1}$ and $X_{2}$ \ are independent and given $(X_{1},X_{2},X_{3}),$ $X_{4}$ only depends on $X_{2}.$ 

\item The \emph{factors} are the conditional densities/ mass function $$
\left\{ p_{di\mathcal{G}}(\boldsymbol{\xi })\triangleq p(x_{i}|\boldsymbol{x}%
_{Q(i)}):i=1,2,\ldots ,m\right\}. $$

\item The \emph{formula} $f_{\mathcal{G}}$ then simply uses a familiar
composition formula of conditional densities both in the idle and intervened system. For
example, when all the components of $\boldsymbol{X}$ are discrete then $f_{%
\mathcal{G}}$ is simply the product of these factors, so 
\begin{eqnarray}
p_{d\mathcal{G}}(\boldsymbol{\xi }) &=&f_{\mathcal{G}}\left( p_{d1\mathcal{G}%
}(\boldsymbol{\xi }),p_{d2\mathcal{G}}(\boldsymbol{\xi }),\ldots ,p_{dm%
\mathcal{G}}(\boldsymbol{\xi })\right) =\prod\limits_{i=i}^{m}p_{d}(x_{i}|%
\boldsymbol{x}_{Q(i)})
\end{eqnarray}%
%Fore example in the BN whose graph is given above this formula is%
%\begin{equation*}
%p_{\mathcal{G}}(\boldsymbol{\xi })=p(x_{1})p(x_{2})p(x_{3}|x_{1},x_{2})p(x_{4}|x_{2})
%\end{equation*}

\item Under the \emph{intervention} $d_{I(\widehat{\boldsymbol{x}})},$ under our notation, \cite{Pearl2000} argues that we
should set 
\begin{eqnarray*}
\mathcal{P}_{\mathcal{G}}^{0}(d_{I}) &=&\left\{ p_{d}(x_{i}|\boldsymbol{x}%
_{Q(i)})=p_{d_{\emptyset }}(x_{i}|\boldsymbol{x}_{Q(i)}):i\notin I\right\} \\
\mathcal{P}_{\mathcal{G}}^{1}(d_{I}) &=&\left\{ p_{d}(x_{i}|\boldsymbol{x}%
_{Q(i)}):i\in I\right\}.
\end{eqnarray*}%
He also argues that, under an obvious embellishment of the processes
represented by $\mathcal{G}$, $p_{d}(x_{i}|\boldsymbol{x}_{Q(i)})$ should
be chosen to be the (degenerate) mass function that sets $P(X_{i}=\widehat{x}%
_{i})=1$. Many other stochastic interventions, where $p_{d}(x_{i}|%
\boldsymbol{x}_{Q(i)})$ are set as alternative randomising densities/mass
functions have since been proposed and it is trivial to check that all such
algebra falls into the general framework above. We simply substitute
different - usually non-degenerate - factors for these types of intervention
in the above taxonomy within our causal algebra.
\end{enumerate}

So this early example of a causal graph is consistent with our general definition. There are a
surprising number of contexts when such a simple map turns out to be valid,
especially within medical and public health domains. However, this causal algebra is not usually appropriate for
military or criminal modelling of causal effects. Other causal technologies
need to be developed to embrace such applications. There are two reasons for
this. Firstly, critical structural information about the underlying idle
process in such contexts is often not efficiently expressed by a BN or
even its dynamic analogues. Secondly, unless the underlying graph of the idle system is embellished,  the impact of any intervention that provokes an intelligent reaction  is typically far more complex than simply substituting one or two factors in the idle system. An additional layer of modeling is required to adapt the original graph of the idle model around an adversary specific causal algebra, which must be defined to model A's predicted intelligent responses. In this paper we show how to use ARA technologies to do this.

Various forms of coloured graph - see for example \cite{hojsgaard2008} - have been developed especially for Gaussian  processes, which can trivially be extended to causal models in an analogous way to the BN. Many of the models of more complex adversarial settings need to be dynamic. The simplest causal algebra for dynamic processes is the Dynamic Bayesian Network (DBN) $\mathcal{G}$ over time steps $t=0,1,\ldots ,T$. The DBN $%
\mathcal{G}$ is equivalent to a BN with graph $\overline{\mathcal{G}}$ whose vertices are time indexed and then inherit the irrelevance statements
implicit in the DBN: see e.g. \cite{Korb} for a precise definition of this
construction. We can now duplicate the algebra defined above on $\overline{%
\mathcal{G}}$ and translate this to $\mathcal{G}$. So such maps fall within
our generic framework, albeit with often a massive set of factors $\mathcal{P}%
_{\mathcal{G}}$ and types of interventions - for example when an intervention
might be applied over a long time period. $\mathcal{G}$ represents a trivial generalisation because it could be seen as a different causal algebra form to $\overline{\mathcal{G}}$. %Because $\mathcal{G}$ rather than $\overline{\mathcal{G}}$ could be seen as a different causal algebra form, it represents a trivial generalisation. 

Another class of graphical model now used for countermeasure strategies for nuclear and food security modelling  under control  \citep{smith2015decision,barons2022} is the multiregression dynamic model \citep{queen1993}. These are essentially a subclass of DBN with latent states and so, using the comment above,  can be seen as a causal graphical model albeit with a transformed semantic. Although less trivially it is also possible to check that the causal algebras proposed for flow graphs of commodity modelling \citep{smith2007causal} and regulatory models \citep{liverani2016bayesian} provide further examples of such general causal  graphical models.  

\subsection{The Chain Event Graph}

It has long been recognised that trees provide an alternative natural graphical framework to express causal hypotheses \citep{robins1986,Shafer}. Over recent decades such classes of model
have been developed into graphical models - containing the class of
probability decision graphs \citep{jaeger2004} - to embed structural hypotheses
and forms of interventions graphically that cannot be directly represented
in a BN. The best development of this class is the staged tree or its
equivalent chain event graph (CEG) represented by a coloured directed acyclic
graph $\mathcal{G}$ \citep{SmithAnderson08, WilkersonSmith21, ThwaitesEtAl2010, CollazzoBook, SmithShenvi18}. This class is particularly useful when the elicited
natural language description explores the different sequences
of events that could unfold. 

For this class of graphical model the atoms of the sample space of the
variable $\xi $ can be read from the root to sink paths of CEG -
shared by its underlying event tree. Interventions directly analogous to
Pearl's ``do'' operations in this new context are of the form $ d_{I(
\widehat{\boldsymbol{\lambda }})} \in \mathbb{D} $. These would force the unit
along edge $\widehat{\lambda }_{i}$ of the underlying graph $\mathcal{G}$ of
the CEG whenever it arrives at $u_{i}$, $i\in I(\widehat{\boldsymbol{\lambda 
}})$,\ $\widehat{\boldsymbol{\lambda }}\triangleq \left\{ \widehat{\lambda }%
_{i}:i\in I\right\} $.

The semantics of the coloured graph $\mathcal{G\in }\mathbb{G}$ of a CEG are
quite different from those of the BN. Nevertheless its structure also
provides a framework for an associated causal algebra  \citep{ThwaitesEtAl2010,CollazzoBook} and so a unique causal map $\Phi $
for each $d_{I(\widehat{\boldsymbol{\lambda }})}\in \mathbb{D}$. The
definition of the four bullets for this case are given below:

\begin{enumerate}
\item We elicit an event tree. Then either by using a model selection algorithm
or by eliciting  information directly, we construct a coloured graph $\mathcal{%
G\in }\mathbb{G}$ of the CEG, in ways described in \cite{CollazzoBook}. If
the events depicted on the event tree have been elicited consistently with
the order $D$ believes they will unfold, then in e.g. \cite{ThwaitesEtAl2010} we argue that - in a wide number of contexts - the processes elicited
that might drive the idle system will continue to hold after the
interventions $d_{I(\widehat{\boldsymbol{\lambda }})}\in \mathbb{D}$. When
this is the case the coloured graph $\mathcal{G}$ of the CEG provides a
valid inferential framework for both the idle and intervened processes.

\item The \emph{factors} $\mathcal{P}_{d\mathcal{G}}$ of this graph $d_{I(%
\widehat{\boldsymbol{\lambda }})}\in \mathbb{D}$ are now the probability
mass functions 
\begin{equation}\left\{ p_{i\mathcal{G}}(\boldsymbol{\xi })\triangleq p(%
\boldsymbol{\lambda }_{i}|u_{i}):i=1,2,\ldots ,m\right\}
\end{equation} where $p(%
\boldsymbol{\lambda }_{i}|u_{i})$ - often called floret probabilities - are the probabilities on the edges $%
\boldsymbol{\lambda }_{i}\triangleq \left\{ \lambda _{ji}:j\in J\right\} $
emanating from differently coloured non-leaf vertices - called \emph{stages}
- $\left\{ u_{i}:i=1,2,\ldots ,m\right\} $ of the graph $\mathcal{G}$ of the
CEG.

\item The \emph{formula} $f_{\mathcal{G}}$ expresses the atomic
probabilities of $\xi $ as the product of the edge probabilities along the
root to leaf path within $\mathcal{G}$ corresponding to each value of $\xi $%
. This implies that: 
\begin{equation}\label{eqn6}
p_{\mathcal{G}}(\xi )=f_{\mathcal{G}}\left( p_{1\mathcal{G}}(\xi ),p_{2%
\mathcal{G}}(\xi ),\ldots ,p_{m\mathcal{G}}(\xi )\right)
=\prod\limits_{\lambda _{ji}\text{ on path }\xi }p(\lambda _{ji}|u_{i}).
\end{equation}

\item In a setting where no intelligent reactions need to be modelled \citep
{ThwaitesEtAl2010}, \cite{CollazzoBook} propose a causal algebra that
asserts that, after the application of the intervention $d_{I(\widehat{%
\boldsymbol{\lambda }})}\in \mathbb{D}$, whenever $i\in I$, we replace the
factors $p(\lambda _{ji}|u_{i})$ in the above formula by $\widehat{p}%
(\lambda _{ji}|u_{i})$ where 
\begin{equation*}
\widehat{p}(\lambda _{ji}|u_{i})=\left\{ 
\begin{array}{cc}
1 & \lambda _{ji}=\widehat{\lambda }_{ji} \\ 
0 & \lambda _{ji}\neq \widehat{\lambda }_{ji}%
\end{array}%
\right.
\end{equation*}%
Thus in our notation $\mathcal{P}_{\mathcal{G}}^{1}(d_{I})=\left\{ \widehat{p%
}(\lambda _{ji}|u_{i}):i\in I\right\} $.
\end{enumerate}

A particularly important example of a class of causal models used
in adversarial setting are Bayesian versions of \emph{micro-simulation models} \citep{birkin2012}.  Like the CEG\textit{\ }these model people as they pass through a number of phases - each represented by a different simulation
component. Formal analyses of these tools have recognised that if such
simulation tools are stochastic then in fact the network of $m$ component
stochastic simulator/emulator models that describes the passages of the
agents is actually a CEG \citep{strong2022bayesian}. So by using the graph $\mathcal{G}$ of this CEG we
can demonstrate that the natural interventions within this class again
provide a general causal algebra analogous to the one above.

As with the BN there are now several dynamic analogues of the CEG each with
its own graphical framework each of which can be used for to build different
causal algebras. For example, the DCEG \citep{Barrclayej} gives us a coloured
version of the well known state space diagram of a Markov process where
positions transform into its states and the markings correspond to block
structures within the state transition matrix. Another graph used within
idle models of plots of people radicalised to extreme violence is the RDCEG, 
\citep{SmithShenvi18,BunninSmith19}. This is a straightforward
elaboration of the DCEG, but deletes from the supporting coloured graph $%
\mathcal{G}$ of the embellished state space diagram the vertex of an
absorbing state and its connecting edges as such states - corresponding to
an adversary aborting a plot. As with the BN, we define the obvious
interventions analogous to interventions in static models by unfolding the
dynamic model into a massive CEG $\overline{\mathcal{G}}$ expressed over all
time steps. We then define the corresponding causal algebra on $\overline{%
\mathcal{G}}$ to define the factors in $\mathcal{P}_{\mathcal{G}}^{1}(d_{I})$
and their change. An example of a CEG used in an adversarial setting later developed into a dynamic analogue is described below.

\section{An Example of a Causal Chain Event Graph}

Here we choose a simple causal graphical model expressed through the semantics of a CEG, which we later use to illustrate how it is possible to build a causal algebra that is capable of modeling the intelligent reactions of an adversary to a proposed intervention. Thus suppose a defender $D$ learns that with probability $p_{Z}$ an adversary $A$'s agent will seek to infiltrate a friendly state undetected through one of 3 ports $B_{1},B_{2},B_{3}$ with respective probabilities
$q_{1},q_{2},q_{3}$. A possible intervention $d_{1}\in\mathbb{D}$ by $D$ is to
instigate much more stringent police checks at the port $B_{1}$. They believe
this would increase the probability of detecting the agent's entry if he
chooses to enter $B_{1}$ from $p_{1}$ to $1$, whilst leaving the detection
probabilities $p_{2}$ and $p_{3}$ at the other two ports unchanged, $p_2<p_3$. Of
strategic interest is to determine the benefit of $d_{1}\in\mathbb{D}_{1}$
over doing nothing, $d_{\emptyset}$.

Here $\mathbb{D}_{1}\triangleq\left\{  d_{\emptyset},d_{1}\right\}  $. Suppose
$D$'s utility $U$ is an indicator about whether or not the agent is detected
crossing into their territory and captured. Then it is easy to check 
\citep{smithbook} that $D$ will choose the decision $d_{1}\in\mathbb{D}_{1}$
over doing nothing when $p_{d_{1}}\geq p$, where $p_{d_{1}},p $ denotes $D$'s
probability that $D$ will detect the agent's entry after enacting $d_{1}$ or
doing nothing respectively.

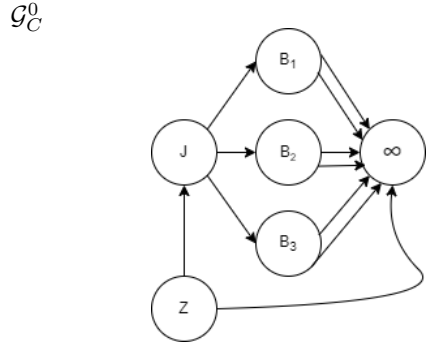
\begin{figure}

    \centering
  \begin{picture}(0,0)
        \put(-50,120){$\mathcal{G}_{C}^{0}$} % Adjust the coordinates (-50,120) as needed 
        \end{picture}
\includegraphics[width=0.3\linewidth]{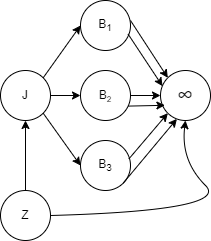}
    \caption{The CEG of an idle incursion model}
    \label{fig:fig1}
\end{figure}

A graph $\mathcal{G}_{C}^{0}$ of the (trivial) CEG representing $D$'s beliefs
about what is happening in this example is given in Fig. \ref{fig:fig1}.

The floret of a vertex $v$ is defined as the sub-tree consisting of $v$, its children, and the edges connecting its children \citep{Freemans08}. 
Therefore, the idle factors of this graph are
the 5 floret probability mass functions \begin{equation}
\left(  p_{Z},1-p_{Z}\right)
,\left(  q_{1},q_{2},q_{3}\right)  ,\left(  p_{1},1-p_{1}\right)  ,\left(
p_{2},1-p_{2}\right)  ,\left(  p_{3},1-p_{3}\right)
\end{equation}respectively
associated with the root vertex and the vertices labelled $B_{i}$, $i=1,2,3.$%

Note that because it is saturated and contains no constraining structural
information other than that logically entailed by the definition of its
florets, $\mathcal{G}_{C}^{0}$ must also hold for any intervened system too.
Denote the corresponding floret probabilities in any intervened systems as
\[
\mathcal{P}_{d}=\left\{  \left(  p_{Z},1-p_{Z}\right)  ,\left(  q_{1d}%
,q_{2d},q_{3d}\right)  ,\left(  p_{1d},1-p_{1d}\right)  ,\left(
p_{2d},1-p_{2d}\right)  ,\left(  p_{3d},1-p_{3d}\right)  \right\}
:d\in\mathbb{D}_{1}
\]
so that in particular $D$'s probability that they will detect the agent's entry
after $d_{1}\in\mathbb{D}_{1}$ will be
\[
p_{d_{1}}\triangleq p_{Z}\left(  q_{1d_{1}}p_{1d_{1}}+q_{2d_{1}}p_{2d_{1}%
}+q_{3d_{1}}p_{3d_{1}}\right).
\]

Denote $D$'s expected utility of doing nothing and doing $d_{1}$ by $\overline
{U}(d_{\emptyset})$ and $\overline{U}(d_{1})$ respectively. It follows that
\[
\overline{U}(d_{\emptyset})=p=p_{Z}\left(  q_{1}p_{1}+q_{2}p_{2}+q_{3}%
p_{3}\right).
\]
To calculate $\overline{U}(d_{1})$ \emph{when the agent is unaware of the
intervention} $d_{1}\in\mathbb{D}_{1}$, $D$ can use $\mathcal{G}_{C}^{0}$ as
an inferential framework for $d_{1}$ as well: using its implied factorisations and
function $f_{\mathcal{G}}$ and simply changing some of the factors within these formulae in the
light of applying $d_{1}\in\mathbb{D}_{1}$.

Thus in this setting the standard CEG inanimate causal rules \citep{ThwaitesEtAl2010} can be applied, where after the intervention $d_{1}$ we simply set
$p_{1d_{1}}^{u}=1$ and copy all other factors in the idle system. So in our
notation set
\begin{align*}
\mathcal{P}_{d_{1}}^{0}  & \triangleq\left\{  \left(  p_{Z},1-p_{Z}\right)
,\left(  q_{1d},q_{2d},q_{3d}\right)  ,\left(  p_{2d},1-p_{2d}\right)
,\left(  p_{3d},1-p_{3d}\right)  \right\}  \\
\mathcal{P}_{d_{1}}^{1}  & \triangleq\left\{  \left(  p_{1d}^{u},1-p_{1d}%
^{u}\right)  =\left(  1,0\right)  \right\}
\end{align*}
implying $D$'s expected utility for enacting $d_{1},$ $\overline{U}%
(d_{1})=p_{d_{1}}^{u}$ where $p_{d_{1}}^{u}$ denotes $D$'s probability of
detecting an unaware agent after intervention $d_{1}$. The CEG formula of the
idle CEG using these new factors on $\mathcal{G}_{C}^{0}$ then gives
\[
p_{d_{1}}^{u}\triangleq p_{Z}\left(  q_{1}1+q_{2}p_{2}+q_{3}p_{3}\right)
\]
and%
\[
\overline{U}(d_{1})-U(d_{\emptyset})=p_{Z}\left(  p_{d_{1}}^{u}-p\right)
=p_{Z}q_{1}\left(  1-p_{1}\right)  >0.
\]
So, as we might expect, if the new regime can be applied with no cost $D$ can safely conclude there is always a benefit in choosing the act
$d_{1}\in\mathbb{D}_{1}$ over doing nothing $d_{\emptyset}$, its efficacy
being increasing in $p_{Z}$ and $q_{1}$ and decreasing in $p_{1}$%
: $\mathcal{G}_{C}^{0}$ and its factorisation are sufficient for this causal analysis. Of course in practice the imposition of the  new regime may well incur an additional cost, but the analysis above is easily adapted to take this into account - for example using standard multi-attribute utility analyses \citep{smithbook}.

\section{Intelligent Causal Algebras}\label{sec:intelligent}

\subsection{Introduction}

All the causal algebras discussed above implicitly assume that $A$ cannot
react to a decision $d\in\mathbb{D}$ so as to soften its impact on them. However,
such responses will be almost inevitable when modelling conflict or the
sub-threshold. In this section, we describe how $D$ can systematically adapt an
original model $\mathcal{G}^{0}$ and embellish its topology into a new graph
$\mathcal{G}$ that is expressive enough to embed a transparent causal algebra that
takes account of a single adversary's potential reactions to $D$'s interventions. The original joint density associated with framework
$\mathcal{G}^{0}$ of the idle system is then simply a margin of the joint
density framed around a more refined graph $\mathcal{G}$. We note that - in
the context when\ $\mathcal{G}$ is a BN - this concept that the idle system is
observed as a margin of a more refined description that can then be exploited
to examine inanimate cause is not new - it is in fact a central concept in
determining whether or not a cause is identifiable \citep{Pearl2000}.
Here we simply extend this concept so that it can be applied to other classes
of graphical causal models for the particular contexts where we can expect a
reaction to an intervention involving variables that are not directly observable from
the idle system. 

The method we propose to embed a model of $A$'s reactions builds on a now well
established form of Bayesian game theory called Adversarial Risk Analysis
(ARA) \citep{banks2015}. Unlike more standard game theoretic approaches,
following \cite{kadane1982} and \cite{Smith1996}, within this genre of game theory, the roles of the two players $D$ and $A$ are not treated symmetrically. Instead the ARA model is built on behalf of just one player -
here the defender $D$. We show below that this approach harmonises well with the general type of graphical causal Bayesian analysis we have described above. It
enables intelligent causal algebras - \ i.e. causal algebras where $D$
models what they believe will be $A$'s reaction $\boldsymbol{r}(d)$ to a
$d\in\mathbb{D}$ when $A$ discovers it - in a way that is both feasibly
realistic to the domain and formally justified.

The key to this embedding is for $D$ to assume a form of ARA analysis where
$D$ assumes that $A$ is rational in the sense that they are a Subjective
Expected Utility (SEU) maximiser. The defender $D$ can then put themselves in
$A$'s shoes to double guess their responses. Two desirable features of a
causal model \citep{hill1965environment} are that the
underlying processes that it describes \ - here by the causal graph
$\mathcal{G}$ and its semantics - apply not only to the circumstances under
study - here the idle system - but also other circumstances. In this
adversarial setting we assume this is true for not only the different decisions
$d\in\mathbb{D}$ that $D$ entertains, but also that other intelligent modelers
share the same belief in the structure expressed by $\mathcal{G}$. A novel
feature of this work is that it is argued below that - within this Bayesian
setting - in choosing a causal $\mathcal{G}$ as a framework of a causal
algebra, $D$ should also strive to ensure that it is plausible that $A$ might
use $\mathcal{G}$ as a framework for their own reasoning as an SEU decision
maker. We argue that when it is compelling for $D$ to do this, the graphical
common knowledge shared by $D$ and $A$ can vastly simplify $D$'s inferences
under an ARA analysis. This helps to scale up such methods to large problems and
provide a transparent framework around which to explain, explore and adapt the
ramifications of a given model.

In order to achieve this we argue that $D$ will often need to embellish their
original causal structural framework expressed by $\mathcal{G}^{0}$ into a new
framework $\mathcal{G}$ that is refined enough to explicitly model $A$'s potential
reactions. A methodology for fleshing out $\mathcal{G}^{0}$ into such a
$\mathcal{G}$ is described and then illustrated below. 

\subsubsection{A Risk Analysis Based on a Causal Graph }

As \cite{kadane1982} pointed out, from a Bayesian perspective, $D$ is free to
choose a model of $A$'s rational reactions however they choose - in particular
the types of rationality behind their reactions $A$ might employ. However, to
develop a suitably generic methodology it is important to give guidance to $D$
about precisely how this might be done. We describe below some assumptions
that if $D$ were to accept would enable them to systematically embed their
beliefs of $A$'s intelligent reactions into their causal inferential framework.

We have argued above that if $D$ believes the underlying process is truly
causal then there will exist a causal graph $\mathcal{G}$ that can be used not
only as a framework for $D$'s probability densities $\left\{  p_{d\mathcal{G}%
}(\boldsymbol{\xi}):d\in\mathbb{D}\right\}$, but also $D$'s beliefs about
what $A$'s beliefs $\left\{  p_{d\mathcal{G}}^{\ast}(\boldsymbol{\xi}%
):d\in\mathbb{D}\right\}  $ might be.

\textbf{Assumption 1: A Causal Common Knowledge Graph}. $D$'s
beliefs about \\ $\left\{  p_{d\mathcal{G}}(\boldsymbol{\xi}):d\in\mathbb{D}%
\right\}  $ can be framed by a causal graph $\mathcal{G}$, which $D$ believes
is also common knowledge \citep{hargreaveheap2004} to $D$ and $A$.

Albeit outside a causal analysis, in previous studies of adversarial BN's
\citep{Smith1996,Allard} and more recently for adversarial
CEGs \citep{Thwaitesgames} it has been argued that for Assumption 1 to be
compelling it must be plausible for $D$ to assume that $A$ will only use
external information available to them when there is any advantage to them to
do so. This enables $D$ to reason about $A$'s processing of structural information
properties embedded in the semantics of the graphs we have explored here. In
fact we have found that by making this assumption $D$ can sometimes reduce the
complexity of a naively chosen graphical framework $\mathcal{G}^{+}$ to a
simpler one $\mathcal{G}^{-}$. We illustrate its use and state this assumption more explicitly below.

\textbf{Assumption 2: Adversarial Parsimony}. $D$ believes that whenever $A$
assigns the same expected utility score to two reactions $r_{1}(d),r_{2}%
(d)\in\mathbb{R}$, $d\in\mathbb{D}$, where $r_{1}(d)$ is an explicit function
of a proper subset of the features of which $D$ is aware and on which
$r_{2}(d)$ is based, then $A$ will prefer $r_{1}(d)$ to $r_{2}(d)$,
$d\in\mathbb{D}$.

In the types of adversarial settings we consider in this paper this is usually
a compelling assumption because were $A$ not to do this then the additional
complexity of $A$'s decision rules might then be exploited by $D$ - as
discussed in \cite{Allard} - in the context where $\mathcal{G}$ is a BN. In this sense D believes that A will choose the most
stable one out of the model perturbations of otherwise equivalent reactions.

Finally, and although not strictly necessary, we follow \cite{banks2015} and recommend that once $D$ has assumed $A$ shares their own inferential framework $\mathcal{G}$, $D$ believes $A$ will assign the factors embellishing this graph as a boundedly rational player
\citep{rios2012}, here a 2 step thinker \citep{alderson2011solving}. This enables $D$ to avoid the
complexities of implications of an infinite regress concerning the factors $A$
might assign to their respective factors. In the context of the domains we
address here this can be expressed as follows.

\textbf{Assumption 3: A 2-step Adversarial Response}. Whilst $D$ believes $A$
might well respond to what $D$ actually does, they believe $A$ will not try to
act in a way to deceive $D$ in order to persuade $D$ to subsequently choose
decisions $d\in\mathbb{D}$ that might be more advantageous to $A$. 

This assumption implicitly assumes that $D$ is one step ahead of $A$: they
take account of $A$'s reasoning as it will arise as a function of the decision
$d\in\mathbb{D}$ that $D$ might commit to. However, in choosing their reaction $D$
believes $A$ will not choose its reactions based on their own beliefs about
$D$'s beliefs $\left\{  p_{d\mathcal{G}}(\boldsymbol{\xi}):d\in\mathbb{D}%
\right\}  $\ beyond those encoded in the topology and colouring of the
shared causal graph $\mathcal{G}$ and its shared factors $\mathcal{P}%
_{\mathcal{G}}^{0}$. 

This is of course a substantive assumption and its applicability to any given
adversarial domain always needs to be checked. For example, an adversary may well
try to influence a defender by disguising their responsive capabilities and so
deceive $D$ into choosing an erroneous  decision $d\in\mathbb{D}$ more
advantageous to $A$. However, in settings like the one we describe in our
running example, where $A$'s acts happen last and after $D$ will have
committed to their action $d\in\mathbb{D}$, it is almost automatic in the
sense that $A$'s act can no longer influence $D$'s acts through such methods.
However, in settings where decisions by $A$ and $D$ are simultaneous and
sequential it can be a heroic assumption. An excellent discussion of these
issues, beyond the scope of this paper, can be found in \cite{stahl1994}.

Embedding accepted dogma from game theory \citep{riosinsua2023}, the new features that need to
be imported by $D$ into $\mathcal{G}^{0}$ so that a new graph $\mathcal{G}$  can capture features about A which will determine their rational responses typically fall into the three categories below.
Depending on the semantics of $\mathcal{G}^{0}$ and $\mathcal{G}$, these
typically can necessitate the introduction of additional vertices, edges and colourings into
$\mathcal{G}$ in ways we illustrate below. These three features are as follows:

\begin{enumerate}
\item Whether or not, and if so when and which decisions $d\in\mathbb{D}$ $A$
\textbf{discovers}: represented by additional vertices labelled by
$\boldsymbol{K}$. 

\item What $D$ believes $A$'s \textbf{intents }might be 
\citep{riosinsua2023}.
Within the Bayesian paradigm these are expressed in $D$'s specification of
$A$'s utility function $U^*$. For simplicity we assume that this intent is
defined in a way that is fixed and invariant to whether no action is taken or
whether any $d\in\mathbb{D}$ is chosen by $D$. So, for example, variables in a BN
describing $U^*$ are founder nodes and events associated with intent are near the
root of a CEG describing the process.

\item What of $A$'s \textbf{resources, capabilities and MO}/ playbook
\citep{BunninSmith19} might help determine how $A$
responds to frustrate $D$'s intervention - expressed by a vector of new
variables $\boldsymbol{C}$. Note that in particular these will constrain the
ways $D$ believes $A$ is capable of reacting - the space of their possible
reactions $\mathbb{R(}d)$: $d\in\mathbb{D}$. Again, at least within any time
slice this set too will be invariant to whether no action is taken or whether any
$d\in\mathbb{D}$ is chosen by $D$.
\end{enumerate}

Once all the features of the problem in the three bullets above are in place, $D$
can then embed into $\mathcal{G}$ how they believe $A$ might actually
\textbf{choose to react} $\boldsymbol{r}\left(  d\right)  \in\mathbb{R(}d)$:
$d\in\mathbb{D}:$ these beliefs here represented by $D$'s random variable
$\boldsymbol{R}\left(  d\right)  :d\in\mathbb{D}$. The factors $p_{di}^{r}$
describing these acts are clearly in $\mathcal{P}_{d\mathcal{G}}^{1}$ and need
to be specified as part of the causal algebra for the particular type of
application considered. There are three steps to construct an intelligent
$\mathcal{G}$ from an initial inanimate causal graph $\mathcal{G}^{0}$:

\begin{enumerate}
\item Embellish the original unintelligent graphical model $\mathcal{G}^{0}$
into an intelligent one $\mathcal{G}^{+}$ in a way that is sufficiently
refined to include all factors embedding the relevant features $\left(U^*,
\boldsymbol{K},\boldsymbol{C}\right)  $. This provides $D$ with a framework
for reasoning about $A$'s (bounded) rationality. 

\item Evoke $A$'s rationality and parsimony to simplify $\mathcal{G}^{+}$ into
a final graph $\mathcal{G}$ for $D$'s inferences about how $A$ might react to
$d\in\mathbb{D}$.

\item Guide $D$ in how to choose appropriately the factors in D's causal model for this graph with the quantitative forms of the
factors - most challengingly those in $\mathcal{P}_{d\mathcal{G}}^{1}$ - so
that $D$ can calculate their own expected utility scores $\left\{  \overline
{U}(d):d\in\mathbb{D}\right\}  $ and hence evaluate the efficacy of each of
their potential interventions.
\end{enumerate}

This protocol is motivated by a rationale that is discussed in great detail in \cite{banks2015}. In summary - conditional on $A$'s utility function $U^*\left(
\boldsymbol{e^*}\right)  $ over $A$'s vector of attributes $\boldsymbol{e^*}$, 
their resources/capabilities/MO $\boldsymbol{C}$ and what $A$ has learned
$\boldsymbol{K}$ about $D$'s chosen intervention $d\in\mathbb{D}$ under the
assumptions above - $D$ will believe $A$ will choose a reaction $\boldsymbol{r}%
^{\ast}(d)\in\mathbb{R(}d)$ to $d\in\mathbb{D}$.\
\[
\boldsymbol{r}^{\ast}(d)\triangleq\arg\max_{r(d)\in\mathbb{R(}d)}\overline
{U}^*\left(  \boldsymbol{r}(d)\right)
\]
where
\[
\overline{U}^*\left(  \boldsymbol{r}(d)\right)  \triangleq\int U^*\left(
\boldsymbol{e^*}\right)  p_{\boldsymbol{r}(d)}^{\ast}(\boldsymbol{e^*}%
|U^*,\boldsymbol{K},\boldsymbol{C})d\boldsymbol{e^*}%
\]
where $p_{\boldsymbol{r}(d)}^{\ast}(\boldsymbol{e^*}|U^*,\boldsymbol{K}%
,\boldsymbol{C})$ is $A$'s density over the attributes  $\boldsymbol{e^*}$ of
$A$'s utility.

Were $D$ able to perfectly predict $p_{\boldsymbol{r}(d)}^{\ast}%
(\boldsymbol{e^*}|U^*,\boldsymbol{K},\boldsymbol{C})$, then because $D$ believes
$A$ is an expected utility maximiser it follows that $D$ will believe $A$'s
reaction $\boldsymbol{r}^{\ast}(d)\in\mathbb{R(}d)$ is known to them for each
contemplated intervention $d\in\mathbb{D}$ and triple $(U^*,\boldsymbol{K}%
,\boldsymbol{C})$. So $D$ can systematically construct the factors $p_{di}%
^{1}\in\mathcal{P}_{d\mathcal{G}\left(  \mathbf{R,},U^*,\boldsymbol{K}%
,\boldsymbol{C}\right)  }^{1}$ needed for their probability model by simply
substituting this perfectly predictable reaction $\boldsymbol{r}^{\ast}(d)\in\mathbb{R(}d)$
into their model of the process. They can then calculate their own expected
utility maximising act $d^{\ast}\in\mathbb{D}$. Note that when $D$ is
uncertain about $\left(U^*,  \boldsymbol{K},\boldsymbol{C}\right)  $, $D$ can
still predict their scores conditional on each value these variables can take
and average over these to calculate their expected scores. Note that  in this rational setting, since $A$ is perfectly predictable to $D$, $A$'s reactions are deterministic functions of the states of this triple and so need not be included in the extended graph $\mathcal{G}_{C}^{+}$.%: see the illustration below. 

Of course, as pointed out in \cite{banks2015} typically $D$ will be uncertain about
the conditional probabilities $p_{\boldsymbol{r}(d)}^{\ast}(\boldsymbol{e^*}%
|U^*,\boldsymbol{K},\boldsymbol{C})$ that $A$ will integrate over to score the
different options that will then determine how they act. However, our general
causal framework provides a simpler setting to the more general setting
described in \cite{banks2015} when $D$ assumes that $A$ will share the same
causal graph $\mathcal{G}$ after $D$ enacts a decision $d\in\mathbb{D}$. This
then greatly simplifies $D$'s task to one of double guessing $A$'s
\emph{factors} that embellish this common knowledge graph $\mathcal{G}$.
Although the elements in $A$'s set of factors - here denoted by $\mathcal{P}%
_{\mathcal{G}}^{\ast}$ - will typically be different from $D$'s corresponding
factors in $\mathcal{P}_{\mathcal{G}}$, they are usually relatively
straightforward for $D$ to estimate. Thus, those associated with natural
phenomena $D$ can plausibly conjecture they will be the same as their own.
Furthermore, others will be degenerate. For example, by definition $A$ will
know for sure $\left(U^*, \boldsymbol{K},\boldsymbol{C}\right)  $ and so their
reactions $\left\{  \boldsymbol{R}(d):d\in\mathbb{D}\right\}  .$ So based on
$D$'s beliefs about $\left( U^*, \boldsymbol{K},\boldsymbol{C}\right)  $ they
are then able to determine a \emph{probability distribution} over $A$'s
reactions as a function of $A$'s factors. Note that the graph $\mathcal{G}$
framing these inferences needs to be sufficiently detailed for $D$ to express
their probabilistic beliefs about $\left(U^*,\boldsymbol{K},\boldsymbol{C}%
\right)  $ as this might apply to \emph{any} $d\in\mathbb{D}$.

The precise way in which these elements can be introduced into a graphical
model obviously depend on the semantics of the graph framing the probability
model as well as what and how $D$ believes $A$ might learn of their
intervention. We illustrate below how it is possible to embellish a causal
graph in this way using the simple incursion model, whose idle structure was given by the graph of the CEG given above.

\subsection{An Incursion Model After a Rational Response}

\paragraph{Constructing a New Graph to Model an Adversary's Intelligence}

\begin{example}
[Incursion continued]Continuing the example above, assume that $D$ believes that
$A$ might discover that $D$ had enacted $d_{1}$ once their agent is in place
and ready to enter one of the ports. Strategic interest continues to be in
determining the benefit of $d_{1}\in\mathbb{D}_{1}$ over doing nothing,
$d_{\emptyset}$.
\end{example}

From the discussion above $D$ needs to construct a more refined graph
$\mathcal{G}_{C}^{+}$ (see Fig. \ref{fig:subfig2}) than $\mathcal{G}_{C}^{0}$ to model $A$'s intelligent
reactions. Here we not only need to embed what $D$ believes $A$ might do in
this setting, but also the choice represented by $J$ of which port his agent will try
to slip through and also how he might react $R$, framed by their capability
$C$ once learning $K$ what $D$ had done (here $d_{1}$). Note in this example
that the only features of $A$'s reaction relevant to $D$ are how $A$ might
adapt their incursion plan - here either to abort their mission or switch
their incursion attempt to ports $B_{2}$ or $B_{3}$. This choice will depend
on what $D$ believes $A$'s intent is - as expressed through $A$'s utility
function $U^*_{1}$ and their capability $C$ to move from outside $B_{1}$ to one of the
other ports. These new critical features are explicitly recognised in the
graph $\mathcal{G}_{C}^{+}$. 
\afterpage{%
\begin{figure}[ht]
    \centering

     \begin{tabular}{cc} % Two columns for two figures in each row

        \begin{subfigure}[b]{0.4\linewidth}
            $\mathcal{G}_{C}^{+}$ \\
            \includegraphics[width=\linewidth]{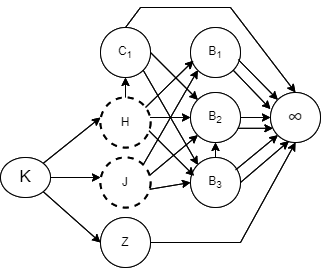}
            \caption{A causal CEG including new nodes to describe knowledge and capability}
            \label{fig:subfig2}
        \end{subfigure}
        &
                
        \begin{subfigure}[b]{0.40\linewidth}
            $\mathcal{G}_{C}^{-}$ \\
             \includegraphics[width=\linewidth]{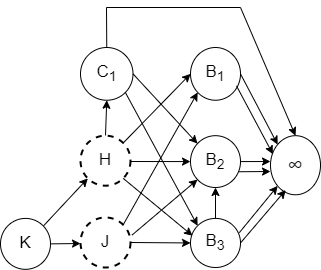}
            \caption{A causal CEG using A's rationality and assumed utility to simplify the inferential frame}
            \label{fig:subfig3}
        \end{subfigure}
        \\
        \vspace{1em}% Adjust the vertical space between rows
      
        \begin{subfigure}[b]{0.4\linewidth}
        $\mathcal{G}_{C}^{-123}$ \\
       
             \includegraphics[width=\linewidth]{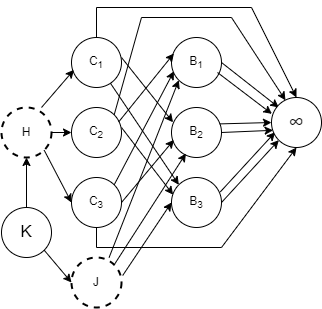}
              \caption{A causal CEG that embraces a larger number of potential intervention}
            
            \label{fig:subfig4}
        \end{subfigure}
    &   
            \begin{subfigure}[b]{0.40\linewidth}
             $\mathcal{G}_{tC}^{+}$\\
             \includegraphics[width=\linewidth]{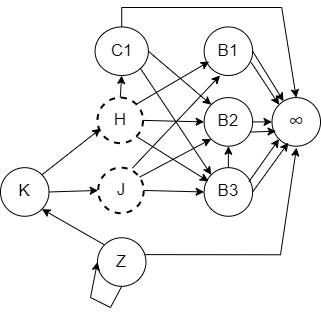} 
                \caption{A Dynamic CEG framework for a different incursion threat}
                \vspace{1em}
                \label{fig:subfig5}
                \end{subfigure}\\
     \end{tabular}

    \vspace{1em} % Adjust the vertical space between the table and the overall caption

    \captionsetup{justification=centering} % Center-align captions
    \caption{Causal Chain Event Graph}
    %\label{fig:overall}
  
\end{figure}

\clearpage%
}
This introduces three new vertices and their
emanating edges: 

\begin{enumerate}
\item A vertex denoted by $K$ whose emanating edges represent whether or not
$D$ believes $A$ discovers $D$'s intervention $d_{1}$ after hearing this
information when the agent has been transported to their planned port of entry.

\item A vertex denoted by $H$ whose edges represent which port the agent might
be positioned to enter when he hears about the intervention $d_{1}$. Note that
here $D$ assumes that $A$ will continue to enter $B_{2}$ or $B_{3}$ after
hearing $d_{1}$ outside one of these ports.

\item A vertex denoted by $C_1$ to represent the event that $A$'s agent is
outside port $B_{1}$ when hearing about $d_{1}$. Its emanating edges are
directed into the port they intelligently react to switch their attempt to
enter given this information. Notice the topology of the CEG asserts they will
always switch to another port in this circumstance if at $C_1$, or abort the
attempted infiltration. There are 3 possible scenarios which challenge the agent's capabilities - not capable of reaching  either $B_{2}$ or $B_{3}$, go to   $B_{2}$ and go to  $B_{3}$. Each of these possible capabilities  is expressed as an emanating edge from $C_1$.
\end{enumerate}

Here the newly introduced floret probabilities are - i.e. the emanating edge probabilities -  $(p_{k},1-p_{k})$ rooted at
the new vertex $K$, the probability the agent hears about $d_{1}$ and floret
probabilities $\left(  q_{1\infty},q_{12},q_{13}\right)  $ rooted at new vertex
$C_1$, $q_{1\infty}$ that the agent at $B_{1}$ chooses or is unable to complete the
mission or $q_{1i}$ that they switch to port $B_{i}$ from $B_{1}$, $i=2,3$ respectively. The story line above ensures that the edge probabilities
$(q_{1},q_{2},q_{3})$ emanating from $H$ are the same as those emanating from
$J$ since $D$ has assumed that $A$ could only learn that $d_{1}$ was put in
place after $A$ had positioned their agent for entry - in CEG terminology $J$
and $H$ are in the same stage and given the same colour in the graph. 

Thus, because $J$ and $H$ are in the same stage, there are now 7 embellishing
probability factors $\mathcal{P}_{\mathcal{G}_{C}^{+}}$ of the new CEG
$\mathcal{G}_{C}^{+}$ emanating from $\left\{  Z,K,J\&H,C_1,B_{1},B_{2}%
,B_{3}\right\}  $ for the idle system
\[
\mathcal{P}_{\mathcal{G}_{C}^{+}}\mathcal{=}\left\{  \left(  p_{Z}%
,1-p_{Z}\right)  ,(0,1),(q_{1},q_{2},q_{3}),(q_{1\infty},q_{12},q_{13})%
,(p_{1},1-p_{1}),(p_{2},1-p_{2}),(p_{3},1-p_{3})\right\}
\]
Obviously in this case, since $d_{1}$ has not been enacted, $p_{k}=0$. Note that $(q_{1\infty},q_{12},q_{13})$ have no effect on the atomic
probabilities of this model using the composition formula of the CEG $\mathcal{G}_{C}^{0}$
because they are always multiplied by zero. Therefore it is trivial to check
that this gives the same formula for the idle system based on $\mathcal{G}%
_{C}^{0}$. In this sense the idle system described by $\mathcal{G}_{C}^{0}$ is
a margin of the idle system described by $\mathcal{G}_{C}^{+}$.

On the other hand, after the intervention $d_{1}$, it is no longer certain
that the now substantive intervention is discovered, so the edge probabilities
$(p_{k},1-p_{k})$ emanating from the vertex $K$ are no longer $\left(
1,0\right)$. Therefore, the new set of factors $\mathcal{P}_{d_{1}\mathcal{G}_{C}^{+}}$ can be partitioned into two sets $\left\{  \mathcal{P}%
_{d_{1}\mathcal{G}_{C}^{+}}^{0},\mathcal{P}_{d_{1}\mathcal{G}_{C}^{+}}%
^{1}\right\}  $ where
\begin{align*}
\mathcal{P}_{d_{1}\mathcal{G}_{C}^{+}}^{0}  & =\left\{  \left(  p_{Z}%
,1-p_{Z}\right)  ,(q_{1},q_{2},q_{3}),(p_{2},1-p_{2}),(p_{3},1-p_{3})\right\}
\\
\mathcal{P}_{d_{1}\mathcal{G}_{C}^{+}}^{1}  & =\left\{  (p_{k},1-p_{k}%
),(q_{1\infty},q_{12},q_{13})\right\}
\end{align*}
where using Equation \eqref{eqn6}, it is easy to check that
$D$'s probability of undetected incursion becomes%
\[
p_{d_{1}}\triangleq p_{Z}p_{k}q_{1}(q_{12}p_{2}+q_{13}p_{3})+(1-p_{k}%
)q_{1}+q_{2}p_{2}+q_{3}p_{3}%
\]
giving that %

\[
\overline{U}(d_{1})-U(d_{\emptyset})=q_{1}\left[  p_{k}(q_{12}p_{2}%
+q_{13}p_{3})+(1-p_{k})-p_{1}\right]
\]
Note now that, in contrast to the case when $A$ cannot intelligently respond,
it is no longer the case that $D$ should always prefer $d_{1}$ to
$d_{\emptyset}$: e.g. when $p_{1},p_{k}\approx1$ and $p_{2},$ $p_{3}\approx0$,
$D$ would judge that it would be better not to intervene.

When performing this sort of extension the first step is to check whether it
is indeed plausible for $D$ to assume that $\mathcal{G}_{C}^{+}$ faithfully
represents their own beliefs and what $A$ might believe. So suppose first that
$D$ believes that $A$ has a utility function $V$ whose only attribute $Y_{1}$
is an indicator on the event that $A$'s agent successfully infiltrates
undetected. Then unusually $U^*_{1}$ is effectively completely known to $D$. The
same argument we used for $D$'s rationality would then lead $D$ to deduce that $A$
will then \emph{minimise} their probability $p^{\ast}$ if their agent is
detected. 

Notice that, having made the assumption about $A$'s utility and that $A$ shares
$\mathcal{G}_{C}^{+}$, $D$ can immediately deduce that $A$ will send the
agent, otherwise $A$'s expected utility will be zero. So they can safely
assume that $p_{Z}=p_{Z}^{\ast}=1$. Therefore the graph $\mathcal{G}_{C}^{+}$
- whilst valid - is over elaborate. It encourages $D$ to introduce unnecessary
features into the analysis. The alternative simpler graph $\mathcal{G}_{C}%
^{-}$ which omits $Z$ is therefore appropriate 

Here
\[
\mathcal{P}_{\mathcal{G}_{C}^{-}}\mathcal{=}\left\{  (0,1),(q_{1},q_{2}%
,q_{3}),(q_{1\infty},q_{12},q_{13}),(p_{1},1-p_{1}),(p_{2},1-p_{2}),(p_{3}%
,1-p_{3})\right\}
\]
and
\begin{align*}
\mathcal{P}_{d_{1}\mathcal{G}_{C}^{-}}^{0}  & =\left\{  (q_{1},q_{2}%
,q_{3}),(p_{2},1-p_{2}),(p_{3},1-p_{3})\right\}  \\
\mathcal{P}_{d_{1}\mathcal{G}_{C}^{-}}^{1}  & =\left\{  (p_{k},1-p_{k}%
),(q_{1\infty},q_{12},q_{13})\right\}
\end{align*}
The hypothesis that $A\ $is SEU with a utility on the given attribute and that
they share a causal graph has already modified $D$'s inferences.

Next, it is necessary to check whether it is plausible for $D$ to entertain a
graphical model with the topology and colouring of the graph $\mathcal{G}_{C}%
^{-}$ given they believe $A$ to be rational. Using the semantics of a CEG the
structural assumptions implied by the graph $\mathcal{G}_{C}^{-}$, see Fig. \ref{fig:subfig3}.

\begin{enumerate}
\item After a possible switch, given whether or not $d_{1}$ is enacted, the
probability that the agent is caught is then unaltered.

\item $A$ will only change their plans if their agent is positioned to enter
port $B_{1}$ when he hears of $d_{1}$.

\item If this happens then $A$ will not attempt to enter $B_{1}$, but enter
either $B_{2}$ or $B_{3}$ or choose/be forced to abort the mission. 
\end{enumerate}

The first assumption is logically implied by $D$'s own beliefs about when $A$
might hear about $d_{1}$. However, the remaining two relate to $D$'s beliefs about
how $A$ will act. Suppose, $D$ believes that $\mathcal{G}_{C}^{-}$ is a
graph of what $D$ believes frames $A$'s probability model given their two
possible decisions. Then the second assumption is automatic if $D$ assumes $A$
is parsimonious. For it is easy to check that whatever $A$'s factors, their
expected utility would not increase through choosing a more complicated
decision to change. 

The third assumption again depends on $D$'s belief that $A$ shares
$\mathcal{G}_{C}^{-}$. For example, if $A$ could infiltrate their agent through
another port $B_{4}$ uncontemplated by $D$ then this assumption would clearly
be violated. However, assume $D$ is confident that this is not the case. Then
this assumption is valid were $D$ to believe that $A$ also believed that after
$d_{1}$ the detection of the agent at port $B_{1}$ was certain. Then, for
any reaction $r$ that included the agent attempting to infiltrate via $B_{1}$
for certain would ensure $A$'s expected utility $\overline{U}^*(r)=0$. So unless
$A$ thinks all other options are hopeless they will choose to try to enter
either $B_{2}$ or $B_{3}.$

Given the topology of the graph $\mathcal{G}_{C}^{-}$ is a valid framework for
expressing $A$'s true beliefs about their attributes $p_{\boldsymbol{r}%
(d)}^{\ast}(\boldsymbol{e^*}|U^*,\boldsymbol{K},\boldsymbol{C})$ and $D$'s beliefs
about these $\widetilde{p}_{\boldsymbol{r}(d)}(\boldsymbol{e^*}|U^*,\boldsymbol{K}%
,\boldsymbol{C})$ - as implied by common knowledge - it
is sufficient for $D$ to consider the respective probabilities
\[
\left\{  p_{k}^{\ast},q_{1}^{\ast},q_{2}^{\ast},q_{3}^{\ast},q_{1\infty}^{\ast
},q_{12}^{\ast},q_{13}^{\ast},p_{1}^{\ast},p_{2}^{\ast},p_{3}^{\ast}\right\}
\]
will be sufficient to construct what $D$ believes $A$'s beliefs are. Although
many of these will be known to $A$, most will be uncertain to $D$. However, $D$
believes that $A$ will try to maximise the value of their infiltrating
undetected: i.e. maximise the function
\[
p_{d_{1}}^{\ast}\triangleq p_{k}^{\ast}q_{1}^{\ast}(q_{12}^{\ast}%
(r)p_{2}^{\ast}+q_{13}^{\ast}(r)p_{3}^{\ast})+(1-p_{k}^{\ast})q_{1}^{\ast
}+q_{2}^{\ast}p_{2}^{\ast}+q_{3}^{\ast}p_{3}^{\ast}%
\]
A key point here is to notice that, through the definition of the variables in
the problem, the only instrument $A$ has is to attempt to get to another port.
This translates into choosing a reaction $r$ to
maximise $q_{12}^{\ast}(r)p_{2}^{\ast}+q_{13}^{\ast}(r)p_{3}^{\ast
}$. The probabilities $(q_{12}^{\ast}(r),q_{13}^{\ast}(r))$ depend on $A$'s
evaluation of arrival capability, which $D$ can place probabilities on but is
unlikely to know precisely. 

$D$ can now  reason as follows. It would be plausible to $D$ to assume that $\left(
p_{2}^{\ast},p_{3}^{\ast}\right)  =\left(  p_{2},p_{3}\right)  $ - after all
these are the current probabilities that $A$ can sample. Note that D here assume A shares their own factor. If D believes this then $A$ will choose
to enter port $B_{2}$ either if $B_{2}$ is the only option, and A believes they cannot reach $B_{3}$, or if A believes they can reach either alternative port and their odds
$\frac{q_{12}^{\ast}(r)}{q_{13}^{\ast}(r)}>\frac{p_{3}}{p_{2}}.$ Such an
assessment would need to be based on $D$'s intelligence about $A$'s
capabilities. This will need to be imported using information not needed in
the idle system - for example information about the ease and the skills of the
agent to travel between the ports. Nevertheless $D$ will have some information
about this.

The probabilities $(q_{12},q_{13})$ then provides $D$'s odds that $A$ switches to
$B_{2}$ and not $B_{3}$, and these provide all the inputs $D$ needs to
assess
\[
\overline{U}(d_{1})-U(d_{\emptyset})=q_{1}\left[  p_{k}(q_{12}p_{2}%
+q_{13}p_{3})+(1-p_{k})-p_{1}\right]
\]
and hence perform their strategic analysis under intelligent cause. Note here that all probabilities used in this equation could be elicited using standard techniques, protocols and code such as those presented in \cite{williams2021}. Of course, the analysis above is the simplest possible. Once we have the graphical model and its factors like this are in place we can, if necessary, bring in the usual Bayesian machinery to bring in our uncertainty about the parameters - using the usual suite of MCMC algorithms. Here this is $D$'s probabilistically expressed uncertainty about conditional probabilities factors (some of which are functions of $A$'s probabilities) and the latent triples $(U^*,\boldsymbol{K},\boldsymbol{C})$. The causal algebra provides us with expected utility values conditional on each iteration of the MCMC algorithm. The implementation of such algorithms within standard ARA models is now well documented and so do not need to be discussed here. However, our point here is that the application of these algorithms greatly simplifies whenever a graphical causal algebra can be constructed that describes the process. Incidentally note that such algorithms can embed $D$'s beliefs that $A$ may not be fully rational, adding uncertainty to $A$'s rational choice $R^*$ - here to their selection of an alternative port when two might be possible. 

\subsubsection{Some Further Points Arising From This Example}

It could be argued that $\mathcal{G}_{C}^{0}$ could still be used by $D$ as a
framework for inference in the example above. After all, subsequent to each
intelligent $d\in{D}_{1}$, $\mathcal{G}_{C}^{0}$ can be calculated as a
margin of $\mathcal{G}_{C}^{-}$ from which the effect margin after
intervention could be calculated. However, then the derived interventional
rules look mysterious. For example, after the intelligent reaction to the
intervention above, the probabilities of an event happening \emph{preceding} an
intervention in the original graph of the idle system $\mathcal{G}_{C}^{0}$ -
there $(q_{1},q_{2},q_{3})$ - will change. Furthermore, \emph{why} and
\emph{how} these change is a function of information not embedded in the
graph. So $\mathcal{G}_{C}^{0}$ $\ $is not sufficiently refined to form any
kind of systematic framework of a causal algebra and to capture the
appropriate time line of events that affect what happens - whilst
$\mathcal{G}_{C}^{+}$ and $\mathcal{G}_{C}^{-}$ are.

Next suppose, confronted with the analysis above, $D$ is still convinced it is
possible that $A$ will abort their plan. Then the arguments above would force
$D$ to conclude that either $A$ is not rational or more likely has
misspecified $A$'s utility. The adversary's preparedness would be entirely
rational if $A$ assigns a positive utility to their agent not being caught.
Thus, within their ARA, $D$ might conjecture that $A$ acts as if they have a
multi-attribute utility $U^*_{12}$ where%
\[
U^*_{12}(e^*_{1},e^*_{2})\triangleq\rho_{v}U^*_{1}(e^*_{1})+\left(  1-\rho_{v}\right)
U^*_{2}(e^*_{2})
\]
where $U^*_{1}(e^*_{1})$ is an indicator on attribute $E_{1}$ - whether
the agent successfully infiltrates as before - and $U^*_{2}(e^*_{2})$ is an indicator
on attribute $E_{2}$ - an event of whether the agent is captured. The parameter $\rho_{v}$, $0<\rho_{v}<1$ defines the criterion weights of this utility \citep{keeney1993,French2010}. It is straightforward
now to rerun the analysis above when it can be checked that $\mathcal{G}%
_{C}^{+}$ defined above, is in fact the appropriate framework to use and cannot
be simplified. The associated equation is then constructed and explained.

Obviously the timing of when $A$ hears about an intervention is again critical
and can influence the topology of the constructed graph. Finally, the necessary
complexity of the graph $\mathcal{G}_{C}$ used to extend $\mathcal{G}_{C}^{0}$
is a function of the contemplated set of interventions $\mathbb{D}$. For example,
suppose $D$ contemplates a larger space of functions $\mathbb{D}_{123}$ that
contains interventions to allocate extra resources at some proper subset of
$B_{1},B_{2},B_{3}$. If these interventions are in place, they will ensure the agent attempts to
infiltrate that port - perhaps randomising across these options. $D$
would then need to embellish $\mathcal{G}_{C}^{-}$ into $\mathcal{G}_{C}^{-123}$
containing new vertices that could describe what the agent would do when
located outside one of the other ports on hearing information about the
tightening of security within these other ports as well so that the efficacy
of the new options could be tested. The graph $\mathcal{G}_{C}^{-123}$ is given in Fig. \ref{fig:subfig4}.

\section{Dynamic Causal Graphs of an Intelligent Adversary}\label{sec:dynamic}

To demonstrate how such methods can scale up, we conclude this section by
introducing a dynamic analogue of the example where an appropriate (now
cyclic) graph the DCEG \citep{Barrclayej} - analogous to a state space diagram
of a semi-Markov process - has a slightly different semantic and the factor
space is much larger. When a defender $D$ intervenes with $d\in\mathbb{D}$
beginning at a time $t$, even when this intervention is initially covert, the
subsequent unfolding of events together with any intelligence sources
available to $A$ will mean that at future time points $D$'s intervention will
be discovered by $A$ (at least partially). So any singular
\emph{interventions} that are planned to be enacted over several time steps
\emph{will only remain covert until a given stopping time when }%
$A$\emph{\ discovers it}. After such a stopping time, $A$ will then be able to
act intelligently. So after that point we will need to apply the predictions
that will embed their intelligent reaction using an appropriately crafted
causal algebra. 

\begin{example}
[Dynamic infiltration]Assume that $D$ learns of a plan by $A$ to infiltrate no
more than one agent on any day $t$ through ports $B_{1},B_{2},B_{3}$ and needs
to quantify the efficacy of a decision $d_{\emptyset t},d_{1t}\in
\mathbb{D}_{1}$ to install the perfect detection system at port $B_{1}$
$s=t,t+1,\ldots,T$ as discussed in the last example. Assume that $D$'s utility
$U$ is linear in the number of agents they detect between day $t$ and day $T$,
$\ t<T$. Assume that $A$ will not learn whether their agent's infiltration
will have been discovered before $T$. The defender $D$ believes that - to make
themselves as unpredictable as possible - $A$ ensures that the event they try
to infiltrate an agent at one of the ports on any given day will be mutually
independent, given the decision they take and whether or not the new system
has been discovered by $A$ by that time $s$.
\end{example}

A DCEG $\mathcal{G}_{tC}^{+}$ of this process is given in Fig. \ref{fig:subfig5}.

Because $J$ and $H$ are in the same stage there are $7(T-t+1)$ factors/ stages
associated with different floret probability vectors emanating from $\left\{
Z,K,J,(H),R1,B_{1},B_{2},B_{3}\right\}  $ over the $(T-t+1)$ time slices. The
factors of the idle system are now respectively
\begin{align*}
\mathcal{P} = \Big\{ & \left( p_{Zs},1-p_{Zs} \right), (0,1), (q_{1s},q_{2s},q_{3s}), \\
& (1-q_{12s},q_{12s}), (p_{1s},1-p_{1s}), (p_{2s},1-p_{2s}), (p_{3s},1-p_{3s}) : s=t,t+1,\ldots,T \Big\}.
\end{align*}
Under the intervention $d_{1}$, in our notation
\begin{align*}
\mathcal{P}_{d_{1}}^{0}  & =\left\{  \left(  p_{Zt},1-p_{Zt}\right)
,(q_{1s},q_{2s},q_{3s}),(p_{2s},1-p_{2s}),(p_{3s},1-p_{3s}):s=t,t+1,\ldots
,T\right\}  \\
\mathcal{P}_{d_{1}}^{R}  & =\left\{  \left(  p_{Zs},1-p_{Zs}\right)
:s=t+1,\ldots,T,(p_{ks},1-p_{ks}),(1-q_{12s},q_{12s}):s=t,t+1,\ldots
,T\right\}.
\end{align*}
However, without additional external information there is no reason for $D$ to
believe that $A$ will change their behaviour. So the probability vectors
\begin{align*}
\Big\{ & (q_{1s},q_{2s},q_{3s}), (1-q_{12s},q_{12s}), \\
& (p_{1s},1-p_{1s}), (p_{2s},1-p_{2s}), (p_{3s},1-p_{3s}), (1-q_{12s},q_{12s}) : s=t,t+1,\ldots, T \Big\}
\end{align*}
will not depend on the time index $s=t,t+1,\ldots,T$. Note that in the dynamic
setting it is possible that $A$ will have discovered $d_{1}$ and then reacts
intelligently to this discovery. So now the probabilities are $\left(
p_{Zs},1-p_{Zs}\right)  ,s=t+1,\ldots,T$ $\in\mathcal{P}_{d_{1}}^{R}$.

So $D$ must now specify $p_{ks}$, the probability that they will learn about
the intervention at time $s=t,t+1,\ldots,T$, and $p_{Zs}$, the probability
they will abort at time $s$, $s=t+1,\ldots,T$. The most naive model for $D$
assumes that $A$'s potential discovery would be totally random to them. Of
course, logically, once $d_{1}$ has been discovered, it remains discovered. So
then $D$ should set
\[
p_{ks}=\sum_{s^{\prime}=t}^{s}p_{k}(1-p_{k})^{s^{\prime}-t}=1-p_{k}^{s-t}.
\]
Using the analogous arguments to the ones above it is possible to check that
the graph $\mathcal{G}_{tC}^{+}$ is justified if $D$ assumes that $A$ has a
utility function that is linear in the number of agents successfully
infiltrating undetected. 

With $D$'s assumed utility it is then easy to calculate that $\overline
{U}(d_{1})-U(d_{\emptyset})$ in this dynamic setting is
\[
\overline{U}(d_{1})-U(d_{\emptyset})=p_{Z}\sum_{s=t}^{T}[p_{k}^{s-t}%
q_{1}\left(  1-p_{1}\right)  +\left(  1-p_{k}^{s-t}\right)]  q_{1}\left[
p_{k}(q_{12}p_{2}+(1-q_{12})p_{3})+(1-p_{k})-p_{1}\right]
\]
So the causal algebra gives us an explicit score for a strategic analysis in
this new dynamic setting in terms of probabilities that can be linked directly
to $D$'s beliefs of $A$'s thought processes - in ways analogous to those
illustrated in the earlier simpler setting. 

\section{Discussion}

In this paper, we have focused on developing a methodological framework for
applying causal graphical inference in an adversarial setting. So we purposely
illustrated the methodology with the simplest possible example that embodies
the application of these ideas where a reactive adversary will try to frustrate
the impacts of any defensive intervention. However, obviously our methods can
also be applied to much more complex environments. So for example,  complex causal analyses and protocols
for adapted graphical forms of ARA are currently being examined for policing terrorist attacks 
orchestrated by hostile nations and for analyses of the possible
exfiltration attacks \citep{BunninSmith19,shenvi2021}. We note that in such cases bespoke graphical frameworks
need to be developed that are demonstrably within the definition of a general
class of causal graphical model given in this paper and where causal
algebras also fully support the types of Bayesian strategic analyses needed in
this setting. We defer the detailed discussion of these more domain specific
causal graphical models to a later paper.

This paper has focused on how to set up prior models for adversarial decision
making. Of course, the appropriate graphical model within a class could be
learned using Bayesian model selection from data, either from analogous
recorded cases or as a particular attack might be unfolding. In particular,
when there is sufficient data available, a best explanatory graph $\mathcal{G}%
$, for example a BN, CEG or one of their dynamic analogues\ can be selected.
The topologies and some of the conditional probability factors shared from the
idle graph of the BN, CEG or alternative graphs can be learned using
standard algorithms provided the data sets are rich enough. 

However, when the intelligent reactions of an adversary need to be modelled it
will hardly ever be the case that\ \emph{all} the factors in $\mathcal{G}$ can
be learned from the idle system. Some will need to be imported from other
data streams. Of course the usual Bayesian inferential framework and subsequent
model selection is then ideal for guiding such selection. We nevertheless note
that for at least some of the factors, because of their novelty or immediacy,
$D$ will need to rely on elicited expert judgements to quantify their model.
Again this is why Bayesian methods are especially valuable in these contexts
where now well developed expert elicitation tools can be applied. Observe that
because there are often systematically missing variables in the extended
intelligent graph, many standard causal discovery tools will fail - although of
course some causal conjectures can sometimes still be investigated for the BN
graphical framework - when large complete observational data sets are available  - using various routine
methods like those discussed in \cite{Spirtesetal1993}. 

One current challenge is to develop bespoke elicitation
methodologies for adversarial domains \citep{riosinsua2023}. We find that directing thought experiments from $D$'s team about $A$'s likely reaction is a very delicate task both about the likely shared graph and the conditional probabilities that D should embellish this with, which properly take account of A's likely reactions. The development for ARA elicitation is
still in its infancy. However, we note that there are many such methods now
developed for various graphical models \citep{Cowell14, Barons2018,WilkersonSmith21,Korb} that could be developed to apply
to adversarial models using the adversarial extensions of these models. The
modularity the graphs induce make such elicitation easier than that
needed more generically for ARA and appear to us a potentially very fruitful
new area of research.

%\section{}\label{}

% \begin{figure} 
% \includegraphics{<eps-file>}% place <eps-file> in ./img  subfolder
% \caption{}
% \label{}
% \end{figure}

% \begin{table} 
% *****************
% \begin{tabular}{lll}
% \end{tabular}
% *****************
% \caption{}
% \label{}
% \end{figure}

%%%%%%%%%%%%%%%%%%%%%%%%%%%%%%%%%%%%%%%%%%%%%%
%% Supplementary Material, if any, should   %%
%% be provided in {supplement} environment  %%
%% with title and short description.        %%
%%%%%%%%%%%%%%%%%%%%%%%%%%%%%%%%%%%%%%%%%%%%%%
%\begin{supplement}
%\stitle{???}
%\sdescription{???.}
%\end{supplement}

%% ** The bibliograhy **
\bibliographystyle{ba}
\bibliography{biblio}

\begin{thebibliography}{42}
\newcommand{\enquote}[1]{``#1''}
\expandafter\ifx\csname natexlab\endcsname\relax\def\natexlab#1{#1}\fi
\expandafter\ifx\csname url\endcsname\relax
  \def\url#1{{\tt #1}}\fi
\expandafter\ifx\csname urlprefix\endcsname\relax\def\urlprefix{URL }\fi
\ifx\endbibitem\undefined \let\endbibitem\relax\fi

\bibitem[{Alderson et~al.(2011)Alderson, Brown, Carlyle, and
  Wood}]{alderson2011solving}
Alderson, D.~L., Brown, G.~G., Carlyle, W.~M., and Wood, R.~K. (2011).
\newblock \enquote{Solving Defender-Attacker-Defender Models for Infrastructure
  Defense.}
\newblock In Wood, R.~K. and Dell, R.~F. (eds.), {\em Operations Research,
  Computing, and Homeland Defense\/}, 28--49. Hanover, MD: INFORMS.
\endbibitem

\bibitem[{Banks et~al.(2015)Banks, Rios, and Ríos~Insua}]{banks2015}
Banks, D., Rios, J., and Ríos~Insua, D. (2015).
\newblock {\em Adversarial Risk Analysis\/}.
\newblock Taylor \& Francis.
\endbibitem

\bibitem[{Barclay et~al.(2015)Barclay, Collazo, Smith, Thwaites, and
  Nicholson}]{Barrclayej}
Barclay, L., Collazo, R., Smith, J., Thwaites, P., and Nicholson, A. (2015).
\newblock \enquote{Dynamic Chain Event Graphs.}
\newblock {\em Electronic Journal of Statistics\/}, 9(2): 2130--2169.
\endbibitem

\bibitem[{Barons et~al.(2022)Barons, Fonseca, Davies, and Smith}]{barons2022}
Barons, M., Fonseca, T., Davies, A., and Smith, J.~Q. (2022).
\newblock \enquote{An Integrating Decision Support System for Addressing Food
  Security in the UK.}
\newblock {\em The Royal Statistical Society Series A\/}, 24.
\endbibitem

\bibitem[{Barons et~al.(2018)Barons, Wright, and Smith}]{Barons2018}
Barons, M., Wright, S., and Smith, J. (2018).
\newblock \enquote{Eliciting Probabilistic Judgments for Integrating Decision
  Support Systems.}
\newblock In Dias, L., Morton, A., and Quigley, J. (eds.), {\em Elicitation:
  The Science and Art of Structuring Judgment\/}, chapter~17, 445--478.
  Springer.
\endbibitem

\bibitem[{Birkin and Wu(2012)}]{birkin2012}
Birkin, M. and Wu, B. (2012).
\newblock \enquote{A Review of Microsimulation and Hybrid Agent-Based
  Approaches.}
\newblock In {\em Agent-Based Models of Geographical Systems\/}, 51--68.
  Springer Netherlands.
\endbibitem

\bibitem[{Bunnin and Smith(2020)}]{BunninSmith19}
Bunnin, F. and Smith, J. (2020).
\newblock \enquote{A Bayesian Hierarchical Model for Criminal Investigations.}
\newblock {\em Bayesian Analysis\/}.
\newblock ArXiv:1907.01894.
\endbibitem

\bibitem[{Collazo et~al.(2018)Collazo, Gorgen, and Smith}]{CollazzoBook}
Collazo, R., Gorgen, C., and Smith, J. (2018).
\newblock {\em Chain Event Graphs\/}.
\newblock Chapman \& Hall.
\endbibitem

\bibitem[{Cowell and Smith(2014)}]{Cowell14}
Cowell, R.~G. and Smith, J.~Q. (2014).
\newblock \enquote{Causal discovery through {MAP} selection of stratified chain
  event graphs.}
\newblock {\em Electronic Journal of Statistics\/}, 8: 965--997.
\endbibitem

\bibitem[{Freeman and Smith(2011)}]{Freemans08}
Freeman, G. and Smith, J. (2011).
\newblock \enquote{Bayesian MAP Selection of Chain Event Graphs.}
\newblock {\em Journal of Multivariate Analysis\/}, 102: 1152--1165.
\endbibitem

\bibitem[{French and Rios~Insua(2010)}]{French2010}
French, S. and Rios~Insua, D. (2010).
\newblock {\em Statistical Decision Theory: Kendall's Library of Statistics
  9\/}.
\newblock Probability \& Mathematical Statistics.
\endbibitem

\bibitem[{Hargreave-Heap and Varoufakis(2004)}]{hargreaveheap2004}
Hargreave-Heap, S. and Varoufakis, Y. (2004).
\newblock {\em Game Theory: A Critical Introduction\/}.
\newblock New York: Routledge.
\endbibitem

\bibitem[{Hill(1965)}]{hill1965environment}
Hill, A.~B. (1965).
\newblock \enquote{The Environment and Disease: Association or Causation?}
\newblock {\em Proceedings of the Royal Society of Medicine\/}, 58(5):
  295--300.
\endbibitem

\bibitem[{Højsgaard and Lauritzen(2008)}]{hojsgaard2008}
Højsgaard, S. and Lauritzen, S.~L. (2008).
\newblock \enquote{Graphical Gaussian models with edge and vertex symmetries.}
\newblock {\em Journal of the Royal Statistical Society: Series B (Statistical
  Methodology)\/}, 70(5): 1005--1027.
\endbibitem

\bibitem[{Jaeger(2004)}]{jaeger2004}
Jaeger, M. (2004).
\newblock \enquote{Probability Decision Graphs – Combining Verification and
  AI Techniques for Probabilistic Inference.}
\newblock {\em International Journal of Uncertainty, Fuzziness, and
  Knowledge-based Systems\/}, 12: 19--42.
\endbibitem

\bibitem[{Kadane and Larkey(1982)}]{kadane1982}
Kadane, J. and Larkey, P. (1982).
\newblock \enquote{Subjective Probability and the Theory of Games.}
\newblock {\em Management Science\/}, 28(2): 113--120.
\endbibitem

\bibitem[{Keeney and Raiffa(1993)}]{keeney1993}
Keeney, R.~L. and Raiffa, H. (1993).
\newblock {\em Decisions with Multiple Objectives: Preferences and Value
  Tradeoffs\/}.
\newblock Cambridge: Cambridge University Press.
\endbibitem

\bibitem[{Korb and Nicholson(2011)}]{Korb}
Korb, K. and Nicholson, A. (2011).
\newblock {\em Bayesian Artificial Intelligence\/}.
\newblock Chapman \& Hall.
\endbibitem

\bibitem[{Liverani and Smith(2016)}]{liverani2016bayesian}
Liverani, S. and Smith, J.~Q. (2016).
\newblock \enquote{Bayesian Selection of Graphical Regulatory Models.}
\newblock {\em International Journal of Approximate Reasoning\/}, 77: 87--104.
\endbibitem

\bibitem[{Naveiro et~al.(2022)Naveiro, Insua, and Camacho}]{naveiro2022}
Naveiro, R., Insua, D.~R., and Camacho, J.~M. (2022).
\newblock \enquote{Augmented probability simulation for adversarial risk
  analysis in general security games.}
\newblock In {\em Proceedings of the 12th International Defense and Homeland
  Security Simulation Workshop (DHSS), 18th International Multidisciplinary
  Modeling \& Simulation Multiconference\/}.
\newblock Awarded with Best Paper Award.
\endbibitem

\bibitem[{Pearl(2000)}]{Pearl2000}
Pearl, J. (2000).
\newblock {\em Causality - Models, Reasoning and Inference\/}.
\newblock Cambridge University Press.
\endbibitem

\bibitem[{Peters et~al.(2016)Peters, Buhlmann, and Meinshaussen}]{PetersRSS}
Peters, J., Buhlmann, N., and Meinshaussen, N. (2016).
\newblock \enquote{Causal Inference by using invariant prediction:
  identification and confidence intervals.}
\newblock {\em JRSSB\/}, 78: 947--1012.
\endbibitem

\bibitem[{Queen and Smith(1993)}]{queen1993}
Queen, C.~M. and Smith, J.~Q. (1993).
\newblock \enquote{Multiregression Dynamic Models.}
\newblock {\em Journal of the Royal Statistical Society Series B (Statistical
  Methodology)\/}, 55(4): 849--870.
\endbibitem

\bibitem[{Rios and Ríos~Insua(2012)}]{rios2012}
Rios, J. and Ríos~Insua, D. (2012).
\newblock \enquote{Adversarial Risk Analysis for Counterterrorism Modeling.}
\newblock {\em Risk Analysis\/}, 32: 894--915.
\endbibitem

\bibitem[{Rios~Insua et~al.(2023)Rios~Insua, Naveiro, Gallego, and
  Poulos}]{riosinsua2023}
Rios~Insua, D., Naveiro, R., Gallego, V., and Poulos, J. (2023).
\newblock \enquote{Adversarial Machine Learning: Bayesian Perspectives.}
\newblock {\em Journal of the American Statistical Association\/}, 1--22.
\endbibitem

\bibitem[{Robins(1986)}]{robins1986}
Robins, J. (1986).
\newblock \enquote{A new approach to causal inference in mortality studies with
  a sustained exposure period—application to control of the healthy worker
  survivor effect.}
\newblock {\em Mathematical Modelling\/}, 7(9--12): 1393--1512.
\endbibitem

\bibitem[{Shafer(1996)}]{Shafer}
Shafer, G.~R. (1996).
\newblock {\em The Art of Causal Conjecture\/}.
\newblock Cambridge, MA: MIT Press.
\endbibitem

\bibitem[{Shenvi et~al.(2021)Shenvi, Bunnin, and Smith}]{shenvi2021}
Shenvi, A., Bunnin, F., and Smith, J. (2021).
\newblock \enquote{A Bayesian Decision Support System for Counteracting
  Activities of Terrorist Groups.}
\newblock {\em Math ArXiv\/}.
\endbibitem

\bibitem[{Smith(1996)}]{Smith1996}
Smith, J. (1996).
\newblock \enquote{Plausible Bayesian Games.}
\newblock In et~al., B. (ed.), {\em Bayesian Statistics 5\/}, 551--560. Oxford
  University Press.
\endbibitem

\bibitem[{Smith(2010)}]{smithbook}
--- (2010).
\newblock {\em Bayesian Decision Analysis: Principles and Practice\/}.
\newblock Cambridge University Press.
\endbibitem

\bibitem[{Smith and Allard(1996)}]{Allard}
Smith, J. and Allard, C. (1996).
\newblock \enquote{Rationality, Conditional Independence and Statistical Models
  of Competition.}
\newblock {\em Computational Learning and Probabilistic Reasoning\/}, 237--256.
\endbibitem

\bibitem[{Smith and Anderson(2008)}]{SmithAnderson08}
Smith, J. and Anderson, P. (2008).
\newblock \enquote{Conditional Independence and Chain Event Graphs.}
\newblock {\em Artificial Intelligence\/}, 172(1): 42--68.
\endbibitem

\bibitem[{Smith and Figueroa(2007)}]{smith2007causal}
Smith, J. and Figueroa, L. (2007).
\newblock \enquote{A Causal Algebra for Dynamic Flow Networks.}
\newblock In Lucas, P., Gamez, J., and Salmeron, A. (eds.), {\em Advances in
  Probabilistic Graphical Models\/}, 39--54. Springer.
\endbibitem

\bibitem[{Smith and Shenvi(2018)}]{SmithShenvi18}
Smith, J. and Shenvi, A. (2018).
\newblock \enquote{Assault Crime Dynamic Chain Event Graphs.}
\newblock Technical report, Warwick Research Report (WRAP).
\endbibitem

\bibitem[{Smith et~al.(2015)Smith, Barons, and Leonelli}]{smith2015decision}
Smith, J.~Q., Barons, M.~J., and Leonelli, M. (2015).
\newblock \enquote{Decision focused inference on Networked Probabilistic
  Systems: with applications to food security.}
\newblock In {\em Proceedings of the Joint Statistical Meeting\/}. Seattle.
\endbibitem

\bibitem[{Spirtes et~al.(1993)Spirtes, Glymour, and Scheines}]{Spirtesetal1993}
Spirtes, P., Glymour, C., and Scheines, R. (1993).
\newblock {\em Causation, Prediction and Search\/}.
\newblock New York: Springer.
\endbibitem

\bibitem[{Stahl and Wilson(1994)}]{stahl1994}
Stahl, D. and Wilson, P. (1994).
\newblock \enquote{Experimental Evidence on Players' Models of Other Players.}
\newblock {\em Journal of Economic Behavior \& Organization\/}, 25: 309--327.
\endbibitem

\bibitem[{Strong et~al.(2022)Strong, McAlpine, and Smith}]{strong2022bayesian}
Strong, P., McAlpine, A., and Smith, J.~Q. (2022).
\newblock \enquote{Towards A Bayesian Analysis of Migration Pathways using
  Chain Event Graphs of Agent Based Models.}
\newblock In {\em Proceedings of the Bayesian Young Statisticians Meeting
  (BaYSM)\/}.
\endbibitem

\bibitem[{Thwaites and Smith(2017)}]{Thwaitesgames}
Thwaites, P. and Smith, J. (2017).
\newblock \enquote{A Graphical Method for Simplifying Bayesian Games.}
\newblock {\em Reliability Engineering and System Safety\/}.
\newblock Online: 12-MAY-2017, DOI: 10.1016/j.ress.2017.05.012.
\endbibitem

\bibitem[{Thwaites et~al.(2010)Thwaites, Smith, and
  Riccomagno}]{ThwaitesEtAl2010}
Thwaites, P., Smith, J., and Riccomagno, E. (2010).
\newblock \enquote{Causal Analysis with Chain Event Graphs.}
\newblock {\em Artificial Intelligence\/}, 174: 889--909.
\endbibitem

\bibitem[{Wilkerson and Smith(2021)}]{WilkersonSmith21}
Wilkerson, R. and Smith, J. (2021).
\newblock \enquote{Customised Structural Elicitation.}
\newblock In Bedford, T., French, S., Hanea, A., and Nane, T. (eds.), {\em
  Expert Judgment in Risk and Decision Analysis\/}. Springer.
\endbibitem

\bibitem[{Williams et~al.(2021)Williams, Wilson, and Wilson}]{williams2021}
Williams, C., Wilson, K., and Wilson, N. (2021).
\newblock \enquote{A Comparison of Prior Elicitation Aggregation Using the
  Classical Method and SHELF.}
\newblock {\em Journal of the Royal Statistical Society Series A: Statistics in
  Society\/}, 184.
\endbibitem

\end{thebibliography}

%\bibliography{bib/sample}
% ** Acknowledgements **
%\begin{acks}[Acknowledgments]
%\end{acks}

\end{document}